\documentstyle[12pt]{article}
\begin{document}
\begin{titlepage}
\mbox{ }
\rightline{SISSA Ref. 139/93/EP}
\rightline{August 1993}
\vspace{1.5cm}
\begin{center}
\begin{Large}
{\bf  Leptonic Decay Constants of Charm and Beauty Mesons in QCD:
      An Update} \footnote{Invited talk at the Third Workshop on
 the Tau-Charm Factory, Marbella, Spain, June 1993. To be published
in the proceedings.}
\end{Large}

\vspace{3cm}

{\large {\bf C. A. Dominguez}}\footnote{On sabbatical leave from Department
of Physics, University of Cape Town, South Africa.}\\[1cm]

Scuola Internazionale Superiore di Studi Avanzati, Trieste, Italy
\end{center}

\vspace{2cm}

\begin{abstract}

An update is given of the determination of leptonic decay constants of
charm and beauty mesons in the framework of relativistic
Hilbert moments and Laplace transform QCD sum rules.
\end{abstract}

\end{titlepage}

\setlength{\baselineskip}{1.5\baselineskip}
\noindent
In this talk I present an update
of relativistic QCD sum rule  estimates of the leptonic decay
constants
\begin{equation}
\langle 0 | A_{\mu} | P(k) \rangle = i \; \sqrt{2} \; f_{P} \;
k_{\mu} ,
\end{equation}
($P = D, D_{s}, B, B_{s}$), so that in this convention $f_{\pi}$ = 93.2 MeV.
The current theoretical status of this problem is rather confusing, due
to the existence of a plethora of predictions not always in mutual
agreement, even if obtained in the same framework.
The major difference between the various determinations to be found
in the literature \cite{Aliev,dp1,rei,nar1,col}
is due to the particular choice of input parameters in the sum rules.
Among these, the values of the heavy quark mass and the asymptotic freedom
threshold impact the most on the
prediction for $f_{P}$. To a lesser extent, sum rule windows or stability
criteria have also an influence on the results. Hence, it is very important
to keep this in mind when comparing different determinations of the same
quantity. It is then not very illuminating to present a table showing
all the various existing predictions.
Instead, I shall present
a single prediction for each $f_{P}$, from Hilbert moments and, separately,
 from Laplace
transform QCD sum rules, using the criterion that the experimental value
of the meson mass should be reproduced by the sum rules, when implemented
by our current best knowledge
of the input parameters. Considerable progress has now been made in
improving the accuracy and reliability of the determinations of
 quark masses \cite{drms},\cite{dpmb}, and  vacuum condensates
\cite{gluon} entering the QCD sum rules.
 The uncertainties in these parameters reflect
in the uncertainty on $f_{P}$, for a given type of sum rule.
The two types of sum rules do not give exactly the same answer for a
given $f_{P}$, although the results are not all that different. A
comparison between the two separate determinations can provide a feeling
for the systematic uncertainties involved in this approach.\\
In order to estimate $f_{P}$ one considers the two-point function
\begin{equation}
\psi_{5}(q^{2}) = i \; \int \; d^{4}x \; e^{iqx} \langle 0 | T
\left( \partial^{\mu} \; A_{\mu}(x) \; \partial^{\nu} \;
A_{\nu}^{\dagger}(0) \right) | 0 \rangle \; ,
\end{equation}
where $\partial^{\mu} \; A_{\mu}(x) = (m_{Q} + m_{q})$ :
$\bar{q}(x) i \gamma_{5} \; Q(x)$: with $q(Q)$ being the light (heavy)
quark field and $m_{q}(m_{Q})$ its corresponding QCD (current) mass. The
function $\psi_{5}(Q^{2}), Q^{2} = -q^{2}$, satisfies a dispersion
relation
\begin{equation}
\psi_{5}(Q^{2}) = \frac {1}{\pi} \; \int \; ds \; \frac {\mbox{Im}
\psi_{5}(s)}{s+Q^{2}} + \mbox{subtractions} ,
\end{equation}
defined up to two subtractions, which can be disposed of by taking at
least two derivatives in (3). In this fashion one obtains the Hilbert
power moments, which at $Q^{2} = 0$ become
\begin{equation}
\varphi^{n}(0) = \frac {(-)^{n+1}}{(n+1)!} \left( \frac {d}{dQ^{2}}
\right)^{n+1} \psi_{5}(Q^{2})|_{Q^{2}=0} = \frac {1}{\pi}
\int^{\infty}_{0} \; \frac {ds}{s^{n+2}} \; \mbox{Im} \psi_{5}(s) \; .
\end{equation}

The point $Q^{2} = 0$ is appropriate for heavy-light quark currents,
to the extent that $\varphi^{n}$ can be computed in perturbative QCD,
adding non-perturbative power corrections which fall off by powers of
the heavy quark mass. These corrections are parametrized by vacuum
expectation values of the quark and gluon fields in the QCD
Lagrangian, and are organized according to their dimension. For
instance, in the limit $m_{q} \rightarrow 0$, well justified for $D_{u,d}$
and $B_{u,d}$ mesons, the perturbative contribution to
$\varphi^{n}(0)$ to order ${\cal{O}}(\alpha_{s})$ is given by \cite{gen}
\begin{equation}
\varphi^{(n)}(0)|_{PT} = \frac {3}{8\pi^{2}} \left( \frac {1}{m_{Q}^{2}}
\right)^{n-1} \; B(n,3) [1 + a^{0}_{n} \; \alpha_{s}] ,
\end{equation}
where $B(x,y)$ is the beta function, $\alpha_{s} \equiv
\alpha_{s}(m_{Q}^{2})$, and $a^{0}_{n}$ are the rational numbers
\begin{eqnarray*}
\frac {3\pi}{4} \; a^{0}_{n} - \frac {\pi^{2}}{6} = 1 - \frac
{2}{n+1} - \frac {6}{n+2} + \sum^{n+2}_{r=1} \; \left[ \frac
{1}{r^{2}} + \left( \frac {3}{2} - \frac {1}{n} \right. \right.
\end{eqnarray*}
\begin{eqnarray}
- \frac {1}{(n+1)} - \frac {1}{(n+2)} + \left. \left.
\frac {3}{(n+3)} \right) \frac {1}{r} \right] \; .
\end{eqnarray}
The non-perturbative part, always in the limit $m_{q} \rightarrow 0$,
becomes \cite{gen}
\begin{eqnarray}
\varphi^{(n)}(0)|_{NP} & = & \frac {-m_{Q}<\bar{q}q>}{m_{Q}^{2n+2}}
\Biggl[ 1 - \frac {<\alpha_{s} G^{2}>}{12\pi m_{Q}<\bar{q}q>} -
\frac {1}{4} (n + 2) (n + 1) \frac {M_{0}^{2}}{m_{Q}^{2}} \Biggr.
\nonumber \\
& - & \Biggl. \frac {4}{81} (n + 2) (n^{2} + 10n + 9) \pi \; \alpha_{s} \;
\rho \; \frac {\langle\bar{q}q\rangle}{m_{Q}^{3}} \Biggr]  ,
\end{eqnarray}

where $\rho$ is a measure of the deviation from the vacuum saturation
approximation of the four-quark condensate ($\rho|_{VS} = 1$).
In the case of the $D_{s}$ and $B_{s}$ mesons, where
the approximation $m_{q} = 0$
should not be made, the full expressions given in \cite{gen} must be used for
$\varphi^{(n)}(0)$. Finally, the hadronic spectral function appearing
on the r.h.s. of (4) is parametrized by the ground state pseudoscalar meson
pole plus a continuum starting at some threshold $s_{0}$. This continuum is
expected to be well approximated by the QCD spectral function, computed in
perturbation theory, provided $s_{0}$ is high enough, i.e.
\begin{equation}
\frac {1} {\pi} \mbox{Im} \psi_{5}(s)|_{HAD} = 2 f_{P}^{2} M_{P}^{4} \;
\delta(s - m_{P}^{2}) + \theta(s - s_{0}) \frac {1}{\pi} \;
\mbox{Im} \psi_{5}(s)|_{PT} \; .
\end{equation}

By taking the ratio of any two consecutive moments one obtains an expression
for $M_{P}^{2}$ as a function of $s_{0}$, the latter being {\it{a -
priori}} unknown. The calculation will be meaningful
provided $M_{P}$ does not depend
strongly on $s_{0}$, i.e. there should be a relatively wide range of values of
$s_{0}$ leading to a value of $M_{P}$ with a reasonably small spread. This is
certainly the case for $D$ and $D_{s}$, where one obtains using the first two
moments $(n = 1,2)$ \cite{dp1}
\begin{equation}
s_{0} = 2M_{D}^{2} - 3 M_{D}^{2} \hspace{1cm} , \hspace{1cm}
M_{D} = 1.85 \pm 0.15 GeV \; ,
\end{equation}
\begin{equation}
s_{0} = 2M_{D_{s}}^{2} - 3 M_{D_{s}}^{2} \hspace{.8cm} , \hspace{1cm}
M_{D_{s}} = 1.9 \pm 0.1 GeV \; ,
\end{equation}
to be compared with the experimental values: $M_{D}|_{EXP} = 1.87$ GeV,
and $M_{D_{s}}|_{EXP} = 1.97$ GeV. With increasing heavy quark mass, the
perturbative contribution increases in importance relative to the
non-perturbative part. Therefore, the stability region in $s_{0}$ becomes
narrower. For instance, for Q=b one finds \cite{dp1} that with
$s_{0} \simeq (1.1 - 1.3) M^{2}_{B}$, the predicted mass is
$M_{B} = 5.2 \pm 0.2$ GeV  ($M_{B}|_{EXP} = 5.27 $ GeV). An additional
{\it confidence criterion} often invoked is that of the hierachy of the
non-perturbative power corrections. An inspection of Eq.(7) shows that
this hierarchy is not respected in the case of Q=c, as the dimension d=5
term can easily become bigger than the d=4 term. In addition, the d=5
contribution could become bigger than the perturbative contribution. In
fact, there are no real solutions for $f_{D}$ and $f_{D_{s}}$, unless
$M_{0}^{2} \leq 0.5 GeV^{2}$ , and $n \leq 2$, as
noticed in  \cite{dp1}. These problems
are absent for Q = b, where the perturbative term dominates the sum rule.
In this case there are real solutions for all values of n, and $M_{0}^{2}$
could be as high as 1 $GeV^{2}$.
In summary, there are certain advantages and shortcomings of the Hilbert
moments, which should be kept in mind when comparing results in this
framework with those from Laplace transform sum rules.\\
The criterion I shall adopt here is to fix  $s_{0}$ in such
a way as to reproduce the experimental value of the pseudoscalar meson
mass, for a given set of input parameters. The latter are chosen as follows:
$m_{c}$ = $1.35 \pm 0.05$ GeV,
 $m_{b}$ =$ 4.72 \pm 0.05$ GeV,
 $m_{s} \simeq 190$ MeV,
 $\Lambda = 200 \pm 100 $ MeV,
 $\rho = 3 \pm 1$,
 $<\alpha_{s} G^{2}> = 0.038 - 0.11$  $GeV^{4}$,
 $<\bar{q} q>|_{c} = - 0.010$ $GeV^{3}$,
 $<\bar{q} q>|_{b} = - 0.014$  $GeV^{3}$,
 and $M_{0}^{2} = 0.5 - 1.0$  $GeV^{2}$, except for Q=c where $M_{0}^{2} = 0.5$
$GeV^{2}$, as mentioned above.
Concerning $f_{D_{s}}$ and $f_{B_{s}}$, previous analyses \cite{dp1,nar1}
have been made keeping $m_{s}\neq 0$ in the perturbative part, but not
in the non - perturbative
expression of the two-point function (2). This
can be improved by keeping $m_{s} \neq 0$ everywhere (details are given
in \cite{dp93}). An additional improvement, in the case of $f_{B_{s}}$
is possible thanks to the recent measurement of the $B_{s}$ mass
\cite{mbsexp}: $M_{B_{s}}|_{EXP} \simeq 5.37$ GeV.\\
The results from these Hilbert moment QCD sum rules
are shown in Tables 1 and 2, for the two extreme values of the corresponding
heavy quark masses. The minimum and
maximum values of the leptonic decay constants are obtained by varying the
input parameters within the limits given above. Of particular importance
are the results for the ratios
$f_{D_{s}}/f_{D}$, and
$f_{B_{s}}/f_{B}$, which are basically independent of the heavy quark mass,
and quite stable against changes in the rest of the input parameters.\\

\begin{table}[phtb]
\noindent
\begin{center}
\caption{Hilbert moment QCD sum rule results for Q=c}
\vspace{1ex}
\begin{tabular}{|lcc|}             \hline
                             & $m_{c}$ = 1.3 GeV & $m_{c}$ = 1.4 GeV\\ \hline
$f_{D}/f_{\pi}$              & 1.29 -- 1.79      & 1.06 -- 1.51     \\
$s_{0}|_{D}$ ($GeV^{2}$)     & 4.5 -- 8.0        & 4.0 -- 6.0       \\
$f_{D_{s}}/f_{\pi}$          & 1.59 -- 2.06      & 1.33 -- 1.74     \\
$s_{0}|_{D_{s}}$ ($GeV^{2}$) & 5.0 -- 7.5        & 4.5 -- 7.0       \\
$f_{D_{s}}/f_{D}$            & 1.23 -- 1.15      & 1.25 -- 1.15     \\ \hline
\end{tabular}
\end{center}
\end{table}

\begin{table}[phtb]
\noindent
\begin{center}
\caption{Hilbert moment QCD sum rule results for Q = b}
\vspace{1ex}
\begin{tabular}{|lcc|}             \hline
                          & $m_{b}$ = 4.67 GeV & $m_{b}$ = 4.77 GeV\\ \hline
$f_{B}/f_{\pi}$              & 1.19 -- 1.40      & 1.02 -- 1.18     \\
$s_{0}|_{B}$ ($GeV^{2}$)     & 32.5 -- 34.0      & 32.0 -- 33.0     \\
$f_{B_{s}}/f_{\pi}$          & 1.48 -- 1.68      & 1.26 -- 1.44     \\
$s_{0}|_{B_{s}}$ ($GeV^{2}$) & 34.0 -- 35.0      & 33.0 -- 34.0     \\
$f_{B_{s}}/f_{B}$            & 1.24 -- 1.20      & 1.24 -- 1.22     \\ \hline
\end{tabular}
\end{center}
\end{table}

It is possible to choose a different kernel in the dispersion relation (3)
and obtain other types of QCD sum rules, e.g. the Laplace sum rules
\newpage
\begin{eqnarray*}
2f_{P}^{2}  M_{P}^{4} exp (- M_{P}^{2}/M^{2}) =
\int^{so}_{m_{Q}^{2}} \; ds \; exp(-s/M^{2}) \frac {1}{\pi} \mbox{Im}
\psi_{5}(s)|_{QCD}
\end{eqnarray*}
\begin{eqnarray*}
+ m_{Q}^{2} exp(-m_{Q}^{2}/M^{2}) \Biggl[
 <( \alpha_{s}/12 \pi) G^{2} - m_{Q} \bar{q} q > -
\frac {1}{4} \frac {m_{Q}}{M^{2}} \left( 1 - \frac {m_{Q}^{2}}{2M^{2}}
\right)
 2 M_{0}^{2}< \bar{q} q  >
\end{eqnarray*}
\begin{eqnarray}
- \frac {1}{6M^{2}} \left( 2 -
\frac {m_{Q}^{2}}{2M^{2}} - \frac {m_{Q}^{4}}{6M^{4}} \right)
 (16/9) \pi \alpha_{s} \rho < \bar{q} q >^{2}
 \Biggr] \; .
\end{eqnarray}

In (11) $m_{s}=0$ is understood (for details of the case $m_{s} \neq 0$ see
 \cite{dp93}). Notice that the sign of the gluon condensate is positive,
contrary to that in \cite{Aliev}  which is incorrect.
 Taking the first derivative with respect to $M^{2}$ in (11)
gives an independent sum rule  which can be used together with (11) in
order to get an expression for the meson mass $M_{P}$, independent of
$f_{P}$. This procedure, which essentially fixes $s_{0}$, is quite
important, i.e. a determination of $f_{P}$ will be reliable provided that
$M_{P}$ comes out right. This point has not been fully appreciated in
some of the existing analyses.\\
It has often been claimed that Laplace transform QCD sum rules are
superior to the Hilbert moments for the determination of $f_{P}$. I do
not quite agree with this claim. In fact, in spite of the exponential
kernel in the dispersion relation, results are rather sensitive to
$s_{0}$. For instance, in the case of Q=c, a 20\% change in $s_{0}$,
around the value which gives the correct meson mass, induces typically
a 10 \% change in that mass as well as in $f_{D}$. For Q=b, the situation
is somewhat more unstable,
 i.e. a given relative variation in $s_{0}$ is accompanied by roughly
the same relative variation of $M_{B}$ and $f_{B}$. On the other hand, for
a fixed value of $s_{0}$, the predicted $M_{D}$ and $M_{B}$ change by
15 \% and 25 \%, respectively, inside the sum rule windows in the Laplace
parameter $M^{2}$. In spite of all this, it is true that the final
uncertainty in $f_{P}$, due to the uncertainties in the input parameters, is
smaller with the Laplace transform QCD sum rules than with the Hilbert
moments. This can be appreciated from Tables 3 and 4, where I present
the results obtained with the Laplace sum rules.However, the two types
of sum rules exhibit different sensitivities to changes in the input
parameters, in addition to having different advantages and shortcomings. For
this reason they should be viewed as complementary methods within the
general framework of QCD sum rules.
\begin{table}[phtb]
\noindent
\begin{center}
\caption{Laplace transform QCD sum rule results for Q = c}
\vspace{1ex}
\begin{tabular}{|lcc|}             \hline
                          & $m_{c}$ = 1.3 GeV & $m_{c}$ = 1.4 GeV\\ \hline
$f_{D}/f_{\pi}$           & 1.30 -- 1.46        & 1.15 -- 1.29   \\
$s_{0}|_{D}$ ($GeV^{2}$)      & 5.5 -- 5.5      & 5.0 -- 5.0     \\
$f_{D_{s}}/f_{\pi}$           & 1.56 -- 1.75    & 1.45 -- 1.64   \\
$s_{0}|_{D_{s}}$ ($GeV^{2}$)  & 6.0 -- 6.0      & 5.5 -- 5.5     \\
$f_{D_{s}}/f_{D}$            & 1.20 -- 1.20    & 1.26 -- 1.27   \\ \hline
\end{tabular}
\end{center}
\end{table}

\begin{table}[phtb]
\noindent
\begin{center}
\caption{Laplace transform QCD sum rule results for Q = b}
\vspace{1ex}
\begin{tabular}{|lcc|}             \hline
                          & $m_{b}$ = 4.67 GeV & $m_{b}$ = 4.77 GeV\\ \hline
$f_{B}/f_{\pi}$           & 1.32 -- 1.40    & 1.14 -- 1.18   \\
$s_{0}|_{B}$ ($GeV^{2}$) & 35.0 -- 35.5    & 34.5 -- 35.0     \\
$f_{B_{s}}/f_{\pi}$       & 1.58 -- 1.68    & 1.38 -- 1.44   \\
$s_{0}|_{B_{s}}$ ($GeV^{2}$)  & 35.5 -- 36.0 & 35.0 -- 35.5     \\
$f_{B_{s}}/f_{B}$        & 1.20 -- 1.20    & 1.21 -- 1.22   \\ \hline
\end{tabular}
\end{center}
\end{table}

Given the fact that results from the Laplace sum rules
have less of a spread than those from the Hilbert moments, one may
be tempted to consider the former as the best determination of the leptonic
decay constants. However, the two methods are not indepenedent, as the
various vacuum condensates enter {\bf both} sum rules, albeit with different
weight factors, and different signs in the case of  d=5 and d=6. Short of
performing a correlation analysis, I feel
one should not discard the results from
the Hilbert moments (nor perform any average from the two methods), but
rather read the absolute minimum and maximum values from   both
sum rules in conjunction. In any case, the predictions for the ratios
$f_{D_{s}}/f_{D}$, and
$f_{B_{s}}/f_{B}$ turn out to be far less dependent on the values of the
heavy quark masses, and the particular sum rule, leading to the accurate
and stable predictions
\begin{equation}
f_{D_{s}}/f_{D} = 1.21 \pm 0.06 ,\\
\end{equation}
\begin{equation}
f_{B_{s}}/f_{B} = 1.22 \pm 0.02 .\\
\end{equation}
These results are in nice agreement with the expectation that the ratio
between (12) and (13) should be close to unity (see e.g. \cite{grin}).\\
A comparison of the results listed in Tables 1 - 4 with predictions from
lattice QCD (see e.g. \cite{mart}) shows reasonable agreement for
$f_{D}$, and $f_{D_{s}}$, but not quite for $f_{B}$, although the
ratio $f_{B_{s}}/f_{B}$ does compare well. In the framework of fully
relativistic QCD sum rules, it is simply not possible to obtain
values of $f_{B}$ bigger than what is shown in Tables 1 - 4, if one
uses the current best values of the input parameters. There is a recent
claim to the contrary by Narison \cite{nar2}, but I have been unable
to reproduce his results, which I believe to be incorrect.\\
It is only
when one considers the infinite quark mass limit, and after resumming
the large logarithms \cite{bagan},
 that one can approach lattice QCD predictions. The price to pay, though,
is a two - loop correction at the 100 \% level in the expression for
$f_{B}^{2}$.

{\bf Acknowledgements}\\
The author wishes to thank the organizers of the Third Workshop on the
Tau-Charm Factory
 for their effort
in making this conference a great success. He also thanks Daniele Amati
for his kind hospitality at SISSA, where this report was written.
This work was supported in
part by the Foundation for Research Development.\\


\begin{thebibliography}{99}
\bibitem{Aliev} T.M. Aliev, V.L. Eletsky, Sov. J. Nucl. Phys. {\bf 38}
            (1983) 936.
\bibitem{dp1} C.A. Dominguez, N. Paver, Phys. Lett. {\bf B197} (1987)
423; {\bf B199} (1987) 596 (E).
\bibitem{rei} L.J. Reinders, Phys. Rev. {\bf D38} (1988) 423.
\bibitem{nar1} S. Narison, Phys. Lett. {\bf B198} (1987) 104.
\bibitem{col} P. Colangelo, G. Nardulli, A. A. Ovchinnikov, N. Paver, Phys.
Lett. {\bf B269} (1991) 201.
\bibitem{drms} C. A. Dominguez, E. de Rafael, Ann. Phys. (N.Y.)
{\bf 174} (1987) 372; C. A. Dominguez, C. van Gend, N. Paver, Phys. Lett.
{\bf B253} (1991) 241.
\bibitem{dpmb} C. A. Dominguez, N. Paver, Phys. Lett. {\bf B253} (1991) 241.
\bibitem{gluon} R. A. Bertlmann {\it et al}, Z. Phys. {\bf C39} (1988) 231;
C. A. Dominguez, J. Sola, Z. Phys. {\bf C40} (1988) 63.
\bibitem{gen} D.J. Broadhurst, S.C. Generalis, Open University Report
No. OUT-4102-8/R (1982); S.C. Generalis, Ph.D. Thesis, Open
University Report No. OUT-4102-13 (1984). See also K. Schilcher,
M.D. Tran, N.F. Nasrallah, Nucl. Phys. {\bf B181} (1981) 91.
\bibitem{dp93} C. A. Dominguez, N. Paver, SISSA Report No. SISSA Ref.
133/93/EP (1993).
\bibitem{mbsexp} ALEPH Collaboration, D. Buskulic {\it et al},
Phys. Lett. {\bf B311} (1993) 425.
\bibitem{grin} B. Grinstein, SSCL Report No. SSCL - 492 (1993).
\bibitem{mart} G. Martinelli, these proceedings.
\bibitem{nar2} S. Narison, Z. Phys. {\bf C55} (1992) 671.
\bibitem{bagan} E. Bagan {{\it et al}}, Phys. Lett. {\bf B278} (1992)
457;
 M. Neubert, Phys. Rev. {\bf D45} (1992) 2451.
\end{thebibliography}
\end{document}